\documentclass[11pt,a4paper]{article}
\usepackage{fullpage}

\usepackage{graphicx}
\usepackage{epsfig,amsmath,amssymb,graphicx,lscape,color}
\usepackage{setspace}
\begin{document}
\doublespace

\title{Calculating groundwater response times for flow in heterogeneous porous media. 
}


\author{Matthew J Simpson$^{1}$ $^\dagger$
}

\maketitle

\noindent
1. School of Mathematical Sciences, Queensland University of Technology (QUT), Brisbane, Australia; +61 (0)731385241; matthew.simpson@qut.edu.au \\

\noindent
\textbf{Keywords}: Groundwater, Saturated groundwater flow, Steady state, Transient, Response time.\\

\newpage

\begin{abstract}
Predicting the amount of time required for a transient groundwater response to take place is a practical question that is of interest in many situations.  This time scale is often called the response time.  In the groundwater hydrology literature there are two main methods used to calculate the response time: (i) both the transient and steady state groundwater flow equations are solved, and the response time is taken to be amount of time required for the transient solution to approach the steady solution within some tolerance; and (ii) simple scaling arguments are adopted.  Certain limitations restrict both of these approaches.  Here we outline a third method, based on the theory of mean action time.  We derive the governing boundary value problem for both the mean and variance of action time for confined flow in two-dimensional heterogeneous porous media.  Importantly, we show that these boundary value problems can be solved using widely available software.  Applying these methods to a test case reveals the advantages of the theory of mean action time relative to standard methods.
\end{abstract}

\newpage

\section*{Introduction}\label{sec:Introduction}
Predicting the duration of time required for a transient groundwater response to effectively reach equilibrium is a very practical question that is of interest in almost any situation where a mathematical model is used to study groundwater flow processes (Walton, 2011; Sophocleous, 2012; Currell et al., 2016). In the groundwater hydrology literature this time scale is often called the \textit{response time} (Townley, 1995).  If the time scale of interest in a groundwater modeling exercise is less than the response time then it is appropriate to use a transient mathematical model to study the groundwater flow process (Haitjema, 2006; Bredehoeft and Durbin, 2009).  In contrast, if the time scale of interest is much greater than the response time scale then it might be sufficient to describe the flow process using a much simpler steady state mathematical model (Haitjema, 2006; Bredehoeft and Durbin, 2009).  Developing techniques that can distinguish between scenarios where steady state models are sufficient is useful because the mathematical and computational techniques required to solve the steady state flow equation are much simpler than those required to solve the transient flow equation (Simpson et al. 2003).

In the groundwater hydrology literature two methods are used to calculate the response time:
\begin{itemize}\itemsep=-1mm
\item Method 1: Both the transient and steady state groundwater flow equations are solved, and the response time is taken to be the amount of time taken for the transient solution to approach the steady solution within some predefined tolerance (e.g. Watson et al., 2010; Lu and Werner, 2013; Rousseau-Gueutin et al. 2013); and,
\item Method 2: Simple scaling arguments are adopted. For example, if flow takes place in a confined aquifer with aquifer diffusivity of $D$, then the response time is taken to be approximately $L^2/D$, where $L$ is a relevant length scale (Haitjema 2006; Sophocleous, 2012).
\end{itemize}
Certain limitations restrict the application of these two standard methods.  The first method can be computationally expensive.  Furthermore, the first method also suffers from the restriction that the estimate of the response time depends on the choice of the tolerance, and it is unclear how this choice of tolerance should be made.  For example, should we specify a tolerance of 10\%, 1\%, or 0.1\%? Frustratingly, the estimate of the response time using this method depends on the choice of tolerance and there is no consensus in the literature about what value of the tolerance ought to be adopted.  The second method does not provide any information about spatial variations in the estimate of response time.  Furthermore, it is not obvious how to apply such a simple scaling argument when the aquifer properties are spatially variable, and the aquifer diffusivity might vary by several orders of magnitude.

Here we present the details of a simple computational method, based on the concept of \textit{mean action time} (McNabb and Wake, 1991; Ellery et al. 2012a; Ellery et al. 2012b).  The particular focus of this work  is to derive the governing boundary value problems for both the mean and variance of action time for confined flow in a two-dimensional heterogeneous aquifer.  Importantly, these boundary value problems can be solved very simply, and easily,  using existing computational algorithms that are widely available within the groundwater modelling community.  Applying these methods to a test case reveals how the approach can be used in a practical situation involving multi-dimensional flow through heterogeneous porous media. The test case also highlights the benefit of using the mean action time framework over the two standard methods. A brief discussion about how the concepts described in this work generalise to other flow conditions, and a more explicit discussion about the computational efficiency of the new method concludes this work.

\section*{Theory}\label{Theory} \label{sec:Theory}
Consider two-dimensional, vertically averaged confined flow through a heterogeneous porous medium. The spatial and temporal evolution of the hydraulic head is governed by (Anderson 2007; Bear 1972),
\begin{equation}
S\dfrac{\partial \phi(x,y,t)}{\partial t} = \dfrac{\partial }{\partial x}\left [T(x,y) \dfrac{\partial \phi(x,y,t)}{\partial x} \right]+ \dfrac{\partial }{\partial y}\left [T(x,y) \dfrac{\partial \phi(x,y,t)}{\partial y} \right] + N(x,y,t), \label{eq:flowequation}
\end{equation}
where $\phi(x,y,t)$ is the spatially and temporally variable hydraulic head, $T(x,y)$ is the spatially variable transmissivity, $S$ is the storage coefficient, $N(x,y,t)$ represents sources and sinks (e.g. areal recharge, local pumping/injection), and $t$ is time. The focus of this work is to analyze a finite measure of the amount of time required for some initial condition,
\begin{equation}
\phi_0(x,y) = \phi(x,y,0), \label{eq:initialcondition}
\end{equation}
to evolve, via Equation (\ref{eq:flowequation}),  to some other steady state condition,
\begin{equation}
\phi_{\infty}(x,y) = \lim_{t \to \infty} \phi(x,y,t). \label{eq:steadycondition}
\end{equation}
Strictly speaking, it takes an infinite amount of time to transition from $\phi_{0}(x,y)$ to $\phi_{\infty}(x,y)$, since $\phi_{\infty}(x,y)$ is associated with the long time limit, $t \to \infty$.  However, this strict mathematical definition is not very useful as a practitioner can never wait for an infinite amount of time to pass before using a simpler steady state model to describe the flow process of interest.  This simple observation motivates us to consider a finite estimate of the amount of time to transition from $\phi_{0}(x,y)$ to $\phi_{\infty}(x,y)$.

To begin we consider two quantities (Simpson et al., 2013),
\begin{align}
F(t; x, y) &= 1 - \left[ \dfrac{\phi(x,y,t) - \phi_{\infty}(x,y)}{\phi_{0}(x,y) - \phi_{\infty}(x,y)} \right],\\
f(t; x, y) &= \dfrac{\partial F(t; x, y)}{\partial t},
\end{align}
which, for transitions where $\phi(x,y,t)$ is monotonic in $t$, means that we can treat $F(t; x, y)$ as a cumulative distribution function and $f(t; x, y)$ as the associated probability density function.  Under these conditions a finite estimate of the time scale associated with the transition can be characterised by analyzing the moments of the distribution.  The first two moments, known as the mean action time and variance of action time, respectively, can be written as
\begin{align}
M(x,y) &= \int_{0}^{\infty} t f(t; x, y) \, \textrm{d} t, \label{eq:MAT} \\
V(x,y) &= \int_{0}^{\infty} \left[t - M(x,y)\right]^2 f(t; x, y) \, \textrm{d}t. \label{eq:VAT}
\end{align}
Using these quantities, a practical definition of response time is $t_r(x,y) = M(x,y) + \sqrt{V(x,y)}$ (Simpson et al. 2013; Jazaei et al. 2016a; Jazaei et al. 2016b), corresponding to the mean of the probability density function plus one standard deviation about the mean.  This simple definition accounts for the two most fundamental properties of the probability density function, namely: (i) the mean of the probability density function; and (ii) the spread about the mean (Ellery et al. 2013).  There are three key benefits to working with this definition of response time:
\begin{enumerate}
\item This framework can be used to define the response time without solving Equation (\ref{eq:flowequation}) for $\phi(x,y,t)$;
\item The framework explicitly describes spatial variations in the response time, $t_r(x,y)$, that are neglected when adopting simple scaling approaches; and
\item To apply this framework in a practical situation where flow takes place in a heterogeneous porous medium, the partial differential equations governing $M(x,y)$ and $V(x,y)$ can be solved numerically using well-established algorithms that are widely available.
\end{enumerate}

We will now briefly explain how to arrive at the relevant governing equations required to solve for $t_r(x,y)$. To arrive at the governing equation for $M(x,y)$ we simplify Equation (\ref{eq:MAT}) using integration by parts, taking care to note that the quantity $\phi(x,y,t)-\phi_{\infty}(x,y)$ decays to zero exponentially fast as $t \to \infty$. Combining the resulting expression with Equation (\ref{eq:flowequation}) suggests that we define two new variables for notational convenience:
\begin{align}
\psi(x,y) &= \phi_{\infty}(x,y) - \phi_{0}(x,y), \label{eq:PSI} \\
\xi(x,y) &= \psi(x,y) M(x,y), \label{eq:xi}
\end{align}
which leads to,
\begin{equation}
0 = \dfrac{\partial }{\partial x}\left [T(x,y) \dfrac{\partial \xi(x,y)}{\partial x} \right]+ \dfrac{\partial }{\partial y}\left [T(x,y) \dfrac{\partial \xi(x,y)}{\partial y} \right] + S\psi(x,y). \label{eq:BVPMAT}
\end{equation}
Equation (\ref{eq:BVPMAT}) is analogous to Poisson's equation for $\xi(x,y)$ and so it can be solved using standard numerical methods (Wang and Anderson, 1982).  Once the numerical solution of $\xi(x,y)$ is known, Equation (\ref{eq:xi}) gives $M(x,y)$.

To arrive at the governing equation for $V(x,y)$ we begin with Equation (\ref{eq:VAT}), and note that two of the three integral expressions can be written in terms of $M(x,y)$.  Since our approach will be to solve for $M(x,y)$ and $V(x,y)$ sequentially, we treat $M(x,y)$ as known and we focus on evaluating the third integral in Equation (\ref{eq:VAT}).  Applying integration by parts, combining the resulting expression with Equation (\ref{eq:flowequation}), and defining a new variable for notational convenience:
\begin{equation}
\pi(x,y) = \dfrac{\psi(x,y)}{2} \left[V(x,y) + M^2(x,y) \right], \label{eq:PI}
\end{equation}
we arrive at,
\begin{equation}
0 = \dfrac{\partial }{\partial x}\left [T(x,y) \dfrac{\partial \pi(x,y)}{\partial x} \right]+ \dfrac{\partial }{\partial y}\left [T(x,y) \dfrac{\partial \pi(x,y)}{\partial y} \right] + S\psi(x,y)M(x,y). \label{eq:BVPVAT}
\end{equation}
If we solve for $M(x,y)$ and then $V(x,y)$ sequentially, Equation (\ref{eq:BVPVAT}) is analogous to Poisson's equation for $\pi(x,y)$.  As before, this equation can be solved using standard numerical methods.  Once the numerical solution of $\pi(x,y)$ is known, we can use Equation (\ref{eq:PI}) to calculate $V(x,y)$, and then the response time is given by $t_r(x,y) = M(x,y) + \sqrt{V(x,y)}$. Of course, in order to solve Equations (\ref{eq:BVPMAT}) and (\ref{eq:BVPVAT}) we must also specify boundary conditions and we will explain how the boundary conditions are deduced in the example calculation in Section \ref{sec:Results}.

\section*{Results and discussion} \label{sec:Results}
Through the use of a simple test case we aim to illustrate the application of mean action time theory to flow through heterogeneous porous media.  We consider confined flow through a rectangular domain (Figure 1a) which has a background transmissivity of 100 m$^2$/day, and two elliptical regions with transmissivity of 10 m$^2$/day and 1000 m$^2$/day. The distribution of $T(x,y)$ is given in Figure 1a.  Therefore, in this problem the transmissivity varies over two orders of magnitude.  We consider regional flow driven by areal recharge.  Accordingly we apply a zero flux boundary condition along the vertical boundary where $x=0$ m, and zero flux boundary conditions along both horizontal boundaries where $y=0$ m and $y=500$ m.  The vertical boundary where $x = 1000$ m is held at $\phi = 51$ m.  Flow in the aquifer is driven by constant recharge, $N=0.001$ m/day.  Solving the steady state analogue of Equation (\ref{eq:flowequation}), which can be written as
\begin{equation}
0 = \dfrac{\partial }{\partial x}\left [T(x,y) \dfrac{\partial \phi(x,y)}{\partial x} \right]+ \dfrac{\partial }{\partial y}\left [T(x,y) \dfrac{\partial \phi(x,y)}{\partial y} \right] + N(x,y,t), \label{eq:steadyflowequation}
\end{equation}
gives a steady solution, $\phi(x,y)$, which we take to be the distribution of hydraulic head before the transition takes place, so that  $\phi_{0}(x,y) = \phi(x,y)$ (Figure 1b).   We consider a transition from $\phi_{0}(x,y)$ to a new steady state, $\phi_{\infty}(x,y)$, which is obtained by solving Equation (\ref{eq:steadyflowequation}) for the same transmissivity distribution, recharge rate and boundary conditions, except that the vertical boundary where $x=1000$ m is instantaneously changed to $\phi = 50$ m.  The distribution of $\phi_{\infty}(x,y)$ is given in Figure 1c.

\begin{figure}
\begin{center}
\includegraphics[width=0.6\textwidth]{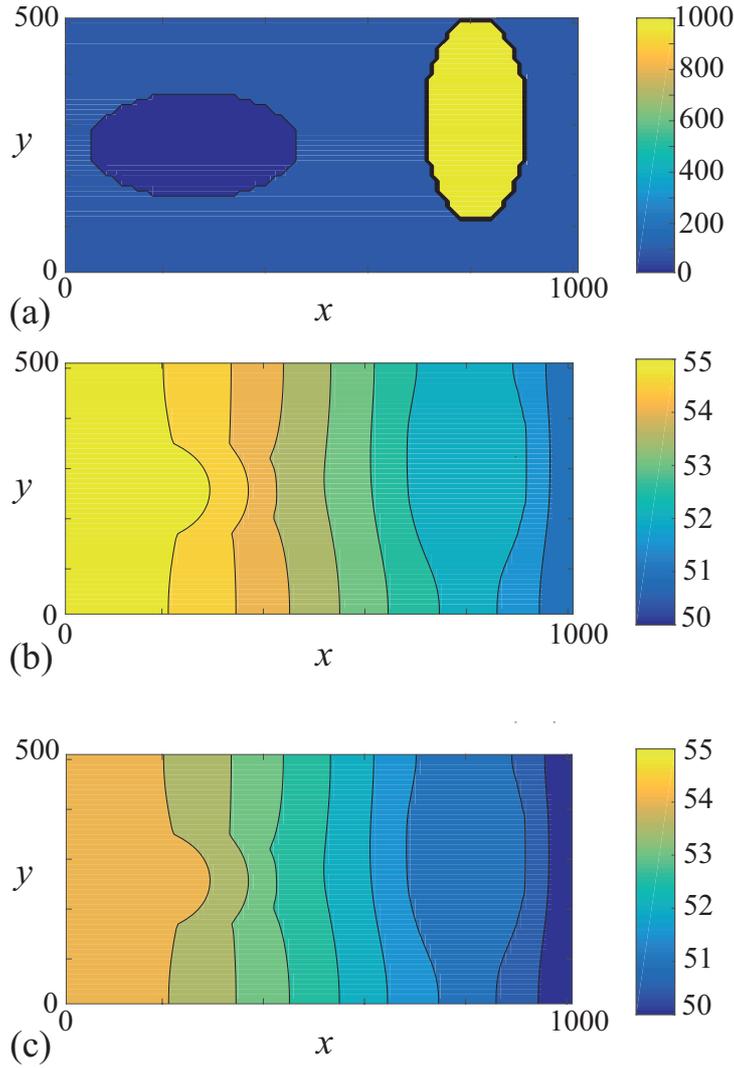}
\caption{ (a) Aquifer transmissivity.  The confined aquifer is characterised by a background transmissivity of 100 m$^2$/day (light blue), with two ellipsoid regions of different transmissivity.  The area enclosed by the ellipsoid with major axis running parallel to the $x$-axis has a lower transmissivity of 10 m$^2$/day (dark blue). The area enclosed by the ellipsoid with major axis running parallel to the $y$-axis has a much higher transmissivity of 1000 m$^2$/day (yellow).  The color bar represents the transmissivity.  Results in (b)-(c) correspond to $\phi_{0}(x,y)$ and $\phi_{\infty}(x,y)$, respectively.  In both (b) and (c) we impose $\partial \phi /\partial y=0$ along both horizontal boundaries and $\partial \phi / \partial x=0$ along the vertical boundary where $x=0$ m.  Results in (b) correspond to setting $\phi = 51$ m along the vertical boundary where $x=1000$ m, whereas results in (c) correspond to setting $\phi = 50$ m along the vertical boundary where $x=1000$ m.  In (b)-(c) the color bar represents the hydraulic head.  Solutions in (b) and (c) correspond to $N(x,y,t)=0.001$ m/day.  Results in (b)-(c) are obtained by solving Equation (\ref{eq:steadyflowequation}) using a finite difference approximation on a square grid, with grid spacing $\delta=10$ m.  The discretized equations are solved iteratively with an absolute convergence tolerance of $10^{-9}$.  All dimensions of length are given in units of metres and all dimensions of time are given in units of days. }
\label{fig:1}       
\end{center}
\end{figure}

To solve Equations (\ref{eq:BVPMAT}) and (\ref{eq:BVPVAT}) we must apply appropriate boundary conditions. Along the vertical boundary where $x = 1000$~m, there is an instantaneous change from $\phi(1000,y)=51$~m to $\phi(1000,y)=50$~m.  Therefore, the appropriate boundary conditions are $M(1000,y)=0$ and $V(1000,y)=0$.  The relevant boundary conditions on the remaining three boundaries requires more consideration.  For example, along the vertical boundary where $x=0$ m, we have $\partial \phi(0,y,t) / \partial x =0$ for $t \ge 0$.  To encode this information as a boundary condition to accompany Equation (\ref{eq:BVPMAT}), we differentiate Equation (\ref{eq:MAT}) with respect to $x$, and evaluate the resulting expression at $x=0$ m.  This gives
\begin{equation}
\dfrac{\partial[M(0,y)\psi(0,y)]}{\partial x} =  \int_{0}^{\infty} \dfrac{\partial \phi(0,y,t)}{\partial x} - \dfrac{\partial \phi_{\infty}(0,y)}{\partial x} \, \textrm{d} t. \label{eq:MATBC}
\end{equation}
The integrand in Equation (\ref{eq:MATBC}) is zero for $t \ge 0$, giving $\partial (M(0,y) \psi(0,y)) / \partial x = 0$, or equivalently, $\partial \xi(0,y) / \partial x=0$ along $x=0$ m.  Similar arguments lead to $\partial \pi(0,y) / \partial x=0$ along $x=0$ m as the appropriate boundary condition for Equation (\ref{eq:BVPVAT}). A similar approach reveals that the boundary conditions along the horizontal boundaries is $\partial \xi / \partial y = 0$ and $\partial \pi / \partial y = 0$, on both horizontal boundaries, where $y=0$ m and $y=500$ m.

Following the approach outlined in Section 2, we solve Equations (\ref{eq:BVPMAT}) and (\ref{eq:BVPVAT}) sequentially, and use Equations (\ref{eq:PSI}) and (\ref{eq:PI}) to evaluate $M(x,y)$ and $V(x,y)$, from which we obtain $t_r(x,t) = M(x,y) + \sqrt{V(x,y)}$. For these calculations we set $S=0.01$. The spatial distribution of $t_r(x,y)$ is given in Figure 2a, illustrating significant spatial variations in the response time for this problem with the regions closer to the boundary at $x=0$ m taking a longer period of time to effectively reach steady state than the regions closer to the vertical boundary at $x=1000$ m.  Furthermore, the spatial patterns of $t_r(x,y)$ show a clear illustration of the effect that the spatial heterogeneity in $T(x,y)$ as the spatial patterns in Figure 1a are echoed in Figure 2a.  In all cases where we present solutions of Equation (\ref{eq:BVPMAT}), Equation (\ref{eq:BVPVAT}) or Equation (\ref{eq:steadyflowequation}), we use an iterative method to solve the linear systems of equations associated with the finite difference discretization approximation.  In all cases the units of the convergence tolerance are the same as the units of the dependent variable.

\begin{figure}
\begin{center}
  \includegraphics[width=0.6\textwidth]{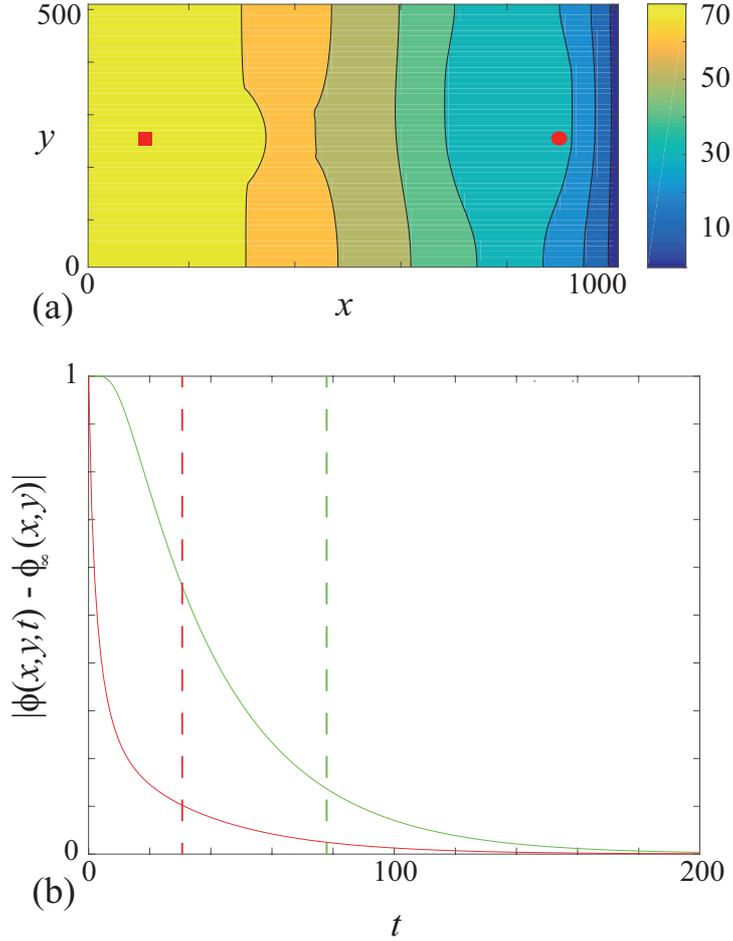}
\caption{ (a) Response time, $t_r(x,y) = M(x,y) + \sqrt{V(x,y)}$ for the transition depicted in Figure 1 from $\phi_0(x,y)$ (Figure 1b) to $\phi_{\infty}(x,y)$ (Figure 1c). The response time calculations correspond to $N(x,y,t)=0.001$ m/day and $S = 0.01$.  The color bar gives $t_r(x,y)$.  To solve for $M(x,y)$ we discretize Equation (\ref{eq:BVPMAT}) on a square finite difference grid, with $\delta = 10$ m.  The resulting equations are solved iteratively with an absolute convergence tolerance of $10^{-9}$.  Similarly, to solve for $V(x,y)$ we discretize Equation (\ref{eq:BVPVAT}) on the same square finite difference grid.  The resulting equations are solved iteratively with an absolute convergence tolerance of $10^{-9}$.  Results in (b) show the time evolution of $|\phi(x,y,t) - \phi_{\infty}(x,y)|$ at two different locations, $(x,y) = (100,250)$ (green) and $(x,y) = (900,250)$ (red).  These locations are indicated in (a) with the red square and red circle, respectively.  The numerical solution of Equation (\ref{eq:flowequation}) is obtained using a central difference approximation on a square finite difference grid, with $\delta = 10$ m.  The coupled system of ordinary differential equations is integrated through time using a second order Runge-Kutta method with constant time steps of duration $\delta t = 0.0002$ days.  Vertical lines in (b) correspond to $t_r(x,y)$ at the two locations, $(x,y) = (100,250)$ (green) and $(x,y) = (900,250)$ (red).   All dimensions of length are given in units of metres and all dimensions of time are given in units of days.}
\label{fig:2}       
 \end{center}
\end{figure}

This example highlights one of the difficulties of using a scaling argument to estimate the response time.  The aquifer diffusivity for this problem varies spatially with $D = T/S$ varying from 1000 m$^2$/day in the low transmissivity region to 100000 m$^2$/day in the high transmissivity region.  If the appropriate length scale for this problem is taken to be the horizontal length scale of the domain, $1000$ m, then a scaling argument implies that the response time lies somewhere between 10 and 1000 days.  The scaling approach is unclear when we consider flow in heterogeneous porous media, and it does not reveal any spatial differences in the response time.  In contrast, the approach based on the theory of mean action time provides a physically-motivated, clearly defined response time that includes both spatial variations and explicitly incorporates the effects of underlying heterogeneity.

An illustration of the value of using the theory of mean action time to estimate the response time for this problem is illustrated in Figure 2b where we use a numerical solution of Equation (\ref{eq:flowequation}) to plot the time evolution of the quantity  $|\phi(x,y,t) - \phi_{\infty}(x,y)|$ at two different locations: $(x,y) = (100,250)$; and, $(x,y) = (900,250)$.  This neatly illustrates the fact that the solution at $(900,250)$ effectively reaches steady state before the solution at $(100,250)$.  Therefore, accounting for spatial variations in the response time could be very practical in this context and these details are neglected by standard scaling arguments.

Another useful result that is worthwhile pointing out is that any existing piece of software that can be used to solve the steady state groundwater flow equation, Equation (\ref{eq:steadyflowequation}), can also be used to solve for $t_r(x,y)$.  For example, Equation (\ref{eq:steadyflowequation}) is identical to Equation (\ref{eq:BVPMAT}) if we make the substitution $\xi = \phi$ and $S\phi = N$.  Furthermore, Equation (\ref{eq:steadyflowequation}) is identical to Equation (\ref{eq:BVPVAT}) if we make the substitution $\pi = \phi$ and $S\phi M = N$.  Therefore, any existing software that can be used to solve Equation (\ref{eq:steadyflowequation}) can also be used to solve for $\xi$ and $\pi$ sequentially, and therefore to compute $t_r(x,y) = M(x,y) + \sqrt{V(x,y)}$.

\section*{Conclusions} \label{sec:Conclusions}
In this work we demonstrate how to derive, and solve the equations governing the mean and variance of action time for a model of confined groundwater flow through heterogeneous porous media.  We also demonstrate, by illustration, how the solutions of these equations can be used to make a useful prediction of the response time.  The spatial variation in the response time are partly driven by spatial variations in the properties of the porous media.

Previous applications of the theory of mean action time to problems from groundwater hydrology  focus on flow through homogeneous porous media (Simpson et al. 2013; Jazaei et al. 2014).  The aim of this methods note is to highlight the practical application of this method by exploring more general scenario including flow in heterogeneous porous media.  An interesting and practical feature of this approach is that the numerical methods required to solve for the mean and variance of action time are widely available, but not used for the purposes of calculating the response time.  For example, any software that can be used to solve the equations governing steady state confined groundwater flow in heterogeneous porous media, such as MODFLOW (Harbaugh, 2005), can also be used to solve for $t_r(x,y) = M(x,y) + \sqrt{V(x,y)}$.

The method presented here, based on the theory of mean action time, requires less computational effort than estimating the response time by computing a numerical solution of the transient flow equation to give $\phi(x,y,t)$.  To calculate the response time using the theory of mean action time, the partial differential equations for $M(x,y)$ (Equation (\ref{eq:BVPMAT})), $V(x,y)$ (Equation (\ref{eq:BVPVAT})), $\phi_{0}(x,y)$ (Equation (\ref{eq:steadyflowequation})) and $\phi_{\infty}(x,y)$ (Equation (\ref{eq:steadyflowequation})) must be solved.   These three differential equations are analogous to Poisson's equation, and can be solved using standard numerical methods.  If these differential equations are discretised on the same spatial mesh, with $\mathcal{N}$ nodes, we can calculate the response time by solving four linear systems, each of size $\mathcal{N} \times \mathcal{N}$.  In contrast, if Equation (\ref{eq:flowequation}) is solved on the same spatial mesh using a standard implicit discretization, then a linear system, of size  $\mathcal{N} \times \mathcal{N}$, must be solved at each time step.   If a transient solution is used to study the response time, the transient solution must be obtained for large $t$, and since the time step must be sufficiently small to control truncation error, this method requires the use of a large number of time steps.  Under these conditions the computational effort in solving the partial differential equations associated with the theory of mean action time is less than the computational effort required to solve Equation (\ref{eq:flowequation}) for large $t$.  Even in situations where Equation (\ref{eq:flowequation}) can be solved for large $t$ with minimal computational effort, it is even less computationally demanding to solve partial differential equations required to evaluate the mean and variance of action time.

Although we explicitly present the governing equations, and their numerical solution, for confined flow through a two-dimensional heterogeneous porous media, the ideas presented here also apply to many relevant related problems.  For example, the methods described here can be generalised to apply to confined three-dimensional flows, as well as two- and three-dimensional flows in anisotropic, heterogeneous porous media.  Furthermore, the methods can also be adapted to provide estimates of the response time for saturated unconfined flow by implementing a linearization assumption (Bear, 1972) and then applying the techniques outlined here to the linearised governing equation. Another possible extension could be to use a slightly different definition of the response time.  Here, in all cases we define the response time to be the mean of $f(t;x,y)$ plus one standard deviation, giving $t_r(x,y) = M(x,y) + \sqrt{V(x,y)}$.  Another possibility is to use the mean plus two standard deviations, or, more generally $t_r(x,y) = M(x,y) + k\sqrt{V(x,y)}$ for $k=1,2,3, \ldots$.  While this alternative definition is physically reasonable, it is less satisfying than simply setting $k=1$ because choosing $k=1$  is the most fundamental definition, and it leads to a reasonable estimate of the response time.  Furthermore, simply setting $k=1$ avoids the need for dealing with the question of how to choose the value of $k$, which, in some sense, is just as arbitrary as determining the tolerance used in Method 1.\\

\end{document}